# Joy Learning: Smartphone Application For Children With Parkinson Disease

Mujahid Rafiq[1], Ibrar Hussain[1], Muhammad Arif[3], Kinza Sardar[1], and Ahsan Humayun[2]

[1]Department of Software Engineering, The University of Lahore, Defence Road Campus, Lahore, Pakistan
[2]Dalian University of Technology, China
[3]The Superior University, Gold Campus, Lahore, Pakistan

**Abstract**
Parkinson's is a Neurologic disorder that not only affects the human body but also their social and personal life. Especially children having the Parkinson's disease come up with infinite difficulties in different areas of life mostly in social interaction, communication, connectedness, and other skills such as thinking, reasoning, learning, remembering. This study gives the solution to learning social skills by using smartphone applications. The children having Parkinson's disease (juvenile) can learn to solve social and common problems by observing real-life situations that cannot be explained properly by instructors. The result shows that the application will enhance their involvement in learning and solving a complex problem.

*Keywords*
*Juvenile, Parkinson, Usefulness, Effective, Social Skills, children with Parkinson*

## 1. Introduction

Parkinson's disease is a severe chronic disorder. After Alzheimer's, It is the second most frequent neurodegenerative disease [1]. According to the research, there are an estimated 1.5 million Americans living with Parkinson's disease and more than 10 million people worldwide [2]. Juvenile Parkinsonism is defined as appearing symptoms of Parkinsonism before the age of 21 years [3]. Parkinson Foundation categorized this disease in children as Young Onset Parkinson's disease (YOPD) and according to them, this occurs very rarely in children but important to address and understand [4]. The children having Parkinson having similar symptoms like Rigidity, Bradykinesia, Tremors in hands, arms legs, Jaw, and face. They have low cognition and mental ability as compared to a normal person [3, 5]. Medical literature is available but mostly tech community is ignoring this society. Only a few applications are available for them mentioned in the related work section but there is a huge gap that needs to fulfill to make them a part of normal society.

The children having Parkinson's feel shy, They avoid socializing, feeling disconnect moreover face the difficulty in learning and dealing with daily life situations like asking for help, solving a problem, asking permission in class and many other related issues. Social interaction is a vital problem for children with Parkinson's disease, mainly when they need memorizing or performing certain tasks [6]. The motive behind this current study is to handle the social situations, improve confidence and enhance knowledge of a child.

In this paper, a smartphone application "Joy learning" introduced for children with Parkinson will help them to learn according to their specialized needs Our focus is on Parkinson children's who avoid going outside, explore new things and are unable to handle altered situations. They usually scared to communicate and interact. Through this application, these limitations can be overcome.

The contribution of work includes providing an effective and useful learning platform for the children having Parkinson's disease. Children can learn different real-life objects, and to deal with real-life problems with accurate examples and situations without any extra effort. This work will also contribute to instructors to instruct the student in a new and easy way.

The poster work is divided into V sections. Section I is about the introduction of work, Section II is on the Related work, III section elaborating the experiment design & methodology we used in this study, Section IV is showing the Analysis & Results and finally, the section V concluded the works and enlightened the future directions of this work

## 2. Related Work

In recent years, there has been an increasing amount of computerized technology use for remedial and educational purpose to serve people with Parkinson's disease. Due to specialized requirements, simple learning applications are not so much use for that kind of child. The most popular and useful applications that are specifically for the learning purpose are discussed in this section.

Kindeo [7], is a private space that provides sharing stories, memories or family knowledge. This application provides a strong understanding of the future generation about their family knowledge already saved here. Cove [8], captures the mood and expresses the way one feels. If one likes specific music/songs, they can add description, picture or save. BeatPanic [9], uses soothing colors and provides positive mantras that help one to calm down and control panic attacks by helping regulate breathing and deviate/ divert your focus from a panic attack. Beats Medical [10], helps improve their speech, motor skills, and mobility. Voice Analyst [11] a self- monitoring app works as an instrument for measuring the pitch and volume of one's voice. This application provides remarkable information regarding speech quality. Parkinson's easy call [12] is useful for dexterity patients for making calls with one touch on the screen. Names and numbers added by the user are then displayed as prominent round buttons. A phone call will be initiated by tapping on a specific button.

The applications presented thus far provide evidence that there is a keen need to make specialized applications for the People having Parkinson's disease to make them a part of the society. Recently a detailed review conducted on applications for Parkinson's disease patients [13].

## 3. Experiment Design & Methodology

The procedures of this study were approved by the ethical body of our department furthermore a formal consent was taken from the doctors and teachers dealing with students under-researched. The interface of the application is properly designed by reading careful design guidelines provided by [14], in their work and then the final design was properly reviewed by 2 Human-Computer Interaction (HCI) experts and 1 doctor and 1 educator that particularly deal with children having Parkinson's.

The images used in application were taken by our self to maintain the familiarity of the context of the situation and to avoid the irrelevant situation that is not present actually. Pictures and videos were taken in natural situations that occurred in schools and the environment.

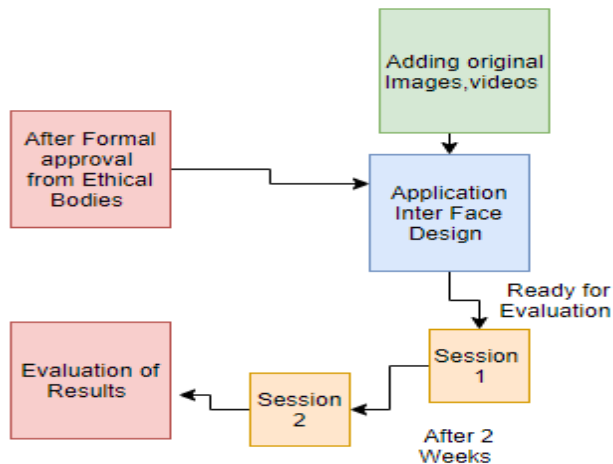

Fig. 1 Overview of Research Methodology

We carefully conducted the experiments on 3 children (2 male and 1 female) having Parkinson's disease and equal intellectual level as told by their instructor. They all belong to Faisalabad, Pakistan. 2 students were aged below 10 and 1 student belongs to the age group from 10-15 years. We assigned tentative names to these children as child A, child B, child C to maintain privacy. Overview of Methodology is showing in Figure 1.

The experiment divided into two session follow-ups, one session is about taking immediate feedback after using the application for the first time and the second session was held after 2 weeks of using application with the help of instructors dealing with these children. In the first session of this experiment, we evaluated the facial expressions, task performance with respect to time, and involvement with respect to interest. We assigned a score out of 100 to these parameters for better analysis, understanding, and measurements. If the score lies in 0-20 % considered worst-performing, 21-40 % bad performing, 41-60 % Average/Not good performing, 61-80 % good performing and 81- 100 % considered as best performance.

Both sessions were performed with the help of 2 instructors and 2 HCI experts. In the second session, improvement in social and common skills was observed. This similar methodology was used in many previous HCI related studies

Design of Application

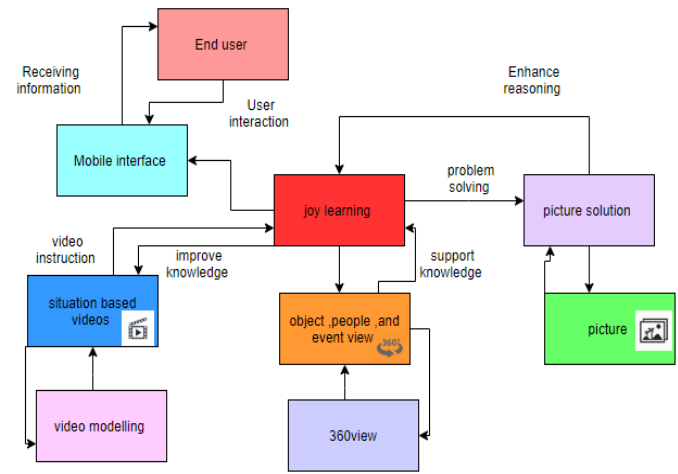

Fig. 2 Joy Learning core components and communication framework.

In figure 2 the core components of "Joy learning" are shown. The application is divided into three parts\modules. We choose the language of the application's interface "English" because existing applications that were used by children and instructors are mostly available in English to avoid leaning effort in the usage of the application. The first part is about "supporting situation" based on video modeling, which has multiple social real-life situations in which application teaches the rules and then provides a similar situation with a different scenario to perform and compare the performance. In this way, we improve social, learning and thinking skills like how to deal with different real-life situations. The second part is "supporting knowledge" that is based on the 360-degree view, characters provide the information of different people, objects and events with an interactive way for those children who are unable to go out and explore objects and people because of bradykinesia and other related symptoms and disorders [3, 15]. Third and the last component is about "supporting solutions" in which we check the children's mental level by providing a picture of a situation based on a specific problem and three options to pick one of the best possible solutions. The selected solution must solve the overall situation. This component also records previous log data of children so that parents and teachers can check their improvement levels in learning. This study primarily evaluates social, mental ability, reasoning, and learning skills by using video modeling, 360-degree view, and specialized pictures.

## 4. Analysis & Results

According to the findings of the first session, child A performed well in comparison with the other 2 children throughout the experiment. Child A had no hesitation and anger on his face with up to 75% positive expressions which are considered well in our case, the task-related performance was up to 66% which is also good and showing 89% involvement in performing tasks that lie in the best category. So the mean of all 3 factors of child A is m=76.6 which means the performance of child A lies in the good category.

The performance of child B was a little bit low in comparison with child A as she was scared while performing the tasks and the score was 36% which was worst in our case. Frustration was showing clearly on her face that's why scored 48% in expressions which are considered bad and 50% score in involvement during performing tasks. The overall mean value of child B= 44.6 which lies in Average/not good category. Evaluation of child C was difficult as he was a quiet child with fewer facial expressions up to 34% that resulted in not gaining any feedback, although he was performing his tasks with up to 60% and showed 70% involvement. The overall mean of child C is m = 54.6 which lies under the Average Performance.

These three children used our application continuously for 2 weeks with the help of the instructor as mentioned in Section III. In the result of the second session, which was based on the measurement of common skills improvements and effects on the social life on the same children, child A showed improvement in learning of common skills up to 83% and learnability on social skills with up to 78% which lies in good and best category and ultimately resulted in better interaction with the society. Child B performed well in the second session with up to 69% of improvement in learning common skills and 60% improvement in the learnability of social life skills. Child C also improved learning with 61% common skills and social skills with up to 54% (which is slightly average/ not good).

The overall results depict that continuous usage of application will definitely effective in the improvement of social as well as common skills for which the application is purely built.

Disclaimer: The results are purely based on user observations. Although the process of data collection is carefully refined and handled with the help of experts there might be chances of minor error and ambiguities which is usually cannot be ignored in human-based experiments.

## 5. Conclusion and Future work

Our purposed smartphone application "Joy learning" is considered useful because as result highlights. It is beneficial in improving the social as well as common skills. Results measure the expressions, performance, and involvement of children while using this application for the very first time. Its interface is designed in a way to handle movement disorders in children with properly suggested guidelines. To make this application more productive, our future aim is to enhance the features of this application, implement it on Virtual Reality platform to give a more real feel about situations and environments. Proper usability experiments will be conducted and will extend the work in a more authentic way. The moreover detailed and extended version of this work will submit in Journal

**Acknowledgment**

We would like to thanks all the persons involved in the survey process and other research-related tasks. Special thanks to Dr. Sonia from Zunnurain Foundation Faisalabad, Pakistan for guidance and opinions related to handling children with Parkinson's disease and help us in conducting tests.

## References

[1] Wirdefeldt, K., Adami, H.-O., Cole, P., Trichopoulos, D., and Mandel, J.: 'Epidemiology and etiology of Parkinson's disease: a review of the evidence', European journal of epidemiology, 2011, 26, (1), pp. 1
[2] https://www.parkinson.org/Understanding Parkinsons/Statistics, accessed June 29 2019
[3] Niemann, N., and Jankovic, J.: 'Juvenile parkinsonism: Differential diagnosis, genetics, and treatment', Parkinsonism & related disorders, 2019

[4] https://www.parkinson.org/Understanding-Parkinsons/What-is-Parkinsons/Young-Onset-Parkinsons, accessed June 29 2019

[5] McNaney, R., Vines, J., Roggen, D., Balaam, M., Zhang, P., Poliakov, I., and Olivier, P.: 'Exploring the acceptability of google glass as an everyday assistive device for people with parkinson's', in Editor (Ed.)^(Eds.): 'Book Exploring the acceptability of google glass as an everyday assistive device for people with parkinson's' (ACM, 2014, edn.), pp. 2551-2554

[6] Schrag, A., Morley, D., Quinn, N., and Jahanshahi, M.: 'Impact of Parkinson's disease on patients' adolescent and adult children', Parkinsonism & related disorders, 2004, 10, (7), pp. 391-397

[7] https://www.parkinsons.org.uk/information-and-support/kindeo, accessed April 21 2019

[8] http://www.cove-app.com/#about, accessed April 16 2019

[9] https://www.parkinsons.org.uk/information-and-support/beat-panic, accessed April 20S 2019

[10] https://www.beatsmedical.com/, accessed April 17 2019

[11] https://www.parkinsons.org.uk/information-and-support/voice-analyst, accessed April 25 2019

[12] http://myhealthapps.net/app/details/396/parkinsons-easycall, accessed June 5 2019

[13] Linares-Del Rey, M., Vela-Desojo, L., and Cano-de la Cuerda, R.: 'Mobile phone applications in Parkinson's disease: A systematic review', Neurología (English Edition), 2018

[14] Nunes, F., Silva, P.A., Cevada, J., Barros, A.C., and Teixeira, L.: 'User interface design guidelines for smartphone applications for people with Parkinson's disease', Universal Access in the Information Society, 2016, 15, (4), pp. 659-679

[15] Jankovic, J.: 'Parkinson's disease: clinical features and diagnosis', Journal of neurology, neurosurgery & psychiatry, 2008, 79, (4), pp. 368-376